# Room Temperature $d^0$ Ferromagnetism in Carbon Doped LaH$_3$: Insights From Density Functional Theory Simulations

Poonam Sharma,[1,] Alok Shukla,[1,*] and Brahmananda Chakraborty,[2, 3*]

[1]*Department of Physics, Indian Institute of Technolog Bombay*, *Mumbai 400076, India*

[2]*High Pressure and Synchrotron Radiation Physics Division,Bhabha Atomic Research Centre, Trombay, Mumbai 400085, India*

[3]*Homi Bhabha National Institute, Mumbai 400094, India*

Email:  shukla@iitb.ac.in, brahma@barc.gov.in

***Abstract*:** Employing the state-of-the-art Density Functional Theory with both GGA and hybrid HSE06 functional along with the incorporation of spin-orbit coupling, we have engineered stable room temperature ferromagnetism in non-magnetic LaH$_3$ through C substitution at octahedral and tetrahedral H sites where the induced magnetic moment is mostly contributed by 2*p* orbital of C atom. It is interesting that the magnetic signature is switched on with an impurity concentration as low as 1.04 at% with a magnetic moment of∼ 1.0μ$_B$ per impurity, where the localized behavior of the 2p states of C along with significant exchange splitting energy can be attributed as the origin of the induced magnetic moment. The verification of the Stoner criterion in the material further confirmed the onset of ferromagnetism in the system, and the computed Curie temperature is found to be well above room temperature. Reduced formation energy and requirement of lower impurity concentration ensure practical feasibility towards a spintronic device where room temperature ferromagnetism is established from the non-magnetic host and the dopant.

## 1. Introduction

Dilute magnetic semiconductors (DMSs) have gained tremendous attention in recent years because of their potential applications in numerous fields such as storage devices, magneto-optical, and spintronics devices [1–3]. For designing the DMS, defects (impurities) are introduced in nonmagnetic (NM) semiconductors with the aim of inducing ferromagnetism (FMs) in them. Traditionally DMSs are synthesized by introducing magnetic impurities such as transition metals (TM) into NM semiconductors because transition metals have unpaired $d$ electrons responsible for magnetism. In addition to charge, the electronic spin degree of freedom also plays a decisive role in these materials, as a result of which they exhibit magnetic as well as semiconducting properties. The advantage of DMS over the semiconductors is that they are nonvolatile, consume less power, enhance the density of integration, and their operation speed is also fast [1,4]. Some of the major applications of DMSs are in spin-polarized light-emitting diode (SPLED), quantum computing devices, spin transistors, lasers, and magneto-optical devices [5–7], to name a few.

By doping Mn into compound semiconductors such as GaAs and InAs, FMs can be obtained [8,9]. However, the Curie temperatures are well below the room temperature for these semiconductors, which renders them less suitable for commercial purposes. For commercial purposes, systems that exhibit FMs against temperature fluctuations and have high Curie temperatures are desired. In general, by exploiting the linear dependence between the Curie temperature and exchange interaction energy, the Curie temperature can be maximized by maximizing the exchange interaction energy for a given system. Further, the interaction strength and, thereby, interaction energy can be manipulated by varying impurity concentrations. If the impurity concentration is very low, the interaction becomes negligible due to the considerable distance between the two adjacent impurity atoms. In contrast, the interaction strength may also become weak for high impurity concentrations due to canceling of the magnetic moments of two closely-placed impurity atoms [10,11]. Therefore, the selection of impurity concentration and thereby obtaining a favorable distance of separation between the impurity atoms that support FMs at room temperature is the crucial step in designing DMS [12].

Even after a decade of research in this area, the origin of induced FMs (i.e., whether induced magnetism is intrinsic or the dopant atoms induce it) and the mechanisms responsible for that are still not understood unambiguously. Although the doping of TM into the host material induces magnetism, its experimental implementation is quite challenging in some cases owing to the clustering of the dopant atoms, which deteriorates the uniformity of the magnetic moment. This, as well as its structural instability, makes the system unsuitable for practical applications [13]. For these reasons, new ways to get the FMs without transition metal doping have been explored. It has been found that magnetism can be induced by doping the NM impurity in the NM semiconductor, called $d^0$ FMs, in which $d$ orbital plays no role in magnetism [14]. These types of new approaches encourage a clearer understanding of the underlying mechanisms responsible for FMs and associated exchange interactions.

The $d^0$ FMs can be induced by cation substitution, anion substitution, impurity adsorption, or creating cation (or anion) vacancies [15–18]. Although all the methods discussed above can yield FMs, the one with the lowest formation energy is the most efficient [17]. Also, it is worth mentioning that in some of the cases, it was observed that the formation energy required for creating a vacancy is substantially more than the cation or anion substitution. Cation or anion substitution was preferred to get the $d^0$ FMs in such cases. Besides this, the impurity concentration and separation between the impurities atoms, as discussed above, are equally responsible for inducing and sustaining (at room temperature) the $d^0$ FMs.

Further, for calculating exchange interaction, the model Hamiltonian has to be constructed according to the electronic structure and the chemical bonding of the system. For magnetic moments, Heisenberg Hamiltonian [19]

$$\mathcal{H} = -\frac{1}{2}\sum_{i \neq j} J_{ij}(\boldsymbol{R}_{ij})\, \boldsymbol{S}_i \cdot \boldsymbol{S}_j \qquad (1)$$

can be used. The exchange interaction constant ($\boldsymbol{J}_{ij}$) is a distance ($\boldsymbol{R}_{ij}$) dependent parameter; $\boldsymbol{S}_i$ and $\boldsymbol{S}_j$ show the spin vectors corresponding to the sites i and j, respectively. One can also use the extensive form of the Heisenberg Hamiltonian by taking other interactions into account as needed [19,20].

Many studies have investigated $d^0$ FMs through cation or anion substitution in different carbides/oxides materials such as ZnO, CaO, HfO$_2$, SiC, and MgO [12,21–24]. Guan et al. studied the FMs in In$_2$O$_3$ using first-principles calculations with the substitution of In by Li/Na/K [25]. Máca et al. stated that in ZrO$_2$, when Zr is substituted with K, FMs is induced in the system with the computed Curie temperature being well above the room temperature, while with Ca substitution, the system remains NM [26]. Chawla et al. experimentally showed that beyond the dopant concentration of 10 at%, FMs disappear in Li-doped ZnO nanorods [12]. Notwithstanding this multitude of studies in carbides/oxides, it is uncommon to find magnetic properties-based studies in hydride. Following these aspects, the present work aims to explore this area, understand the underlying mechanism responsible for FMs, and examine the possibility of the persistence of induced FMs at room temperature on C-doped $LaH_3$. Our finding also encourages for considering hydride classes and exploring them for FMs study.

In our investigations, we show that the C substitution in NM $LaH_3$ introduces a defect band that exhibits large spin splitting near the Fermi level located almost at the middle of the defects band attributing to substantial ferromagnetic (FM) exchange interactions in the system. Moreover, for spontaneous magnetization defect band needs to be adjusted so that the energy gained from the exchange interactions should overcome the kinetic energy loss according to the band picture model in solid. Based on the above fact, we satisfy the stoner criterion to confirm the emergence of FMs in the system [27]. From formation energy calculations, we show that H substitutional doping is the more favorable channel for the entire range of H chemical potential.

## 2. Computational details

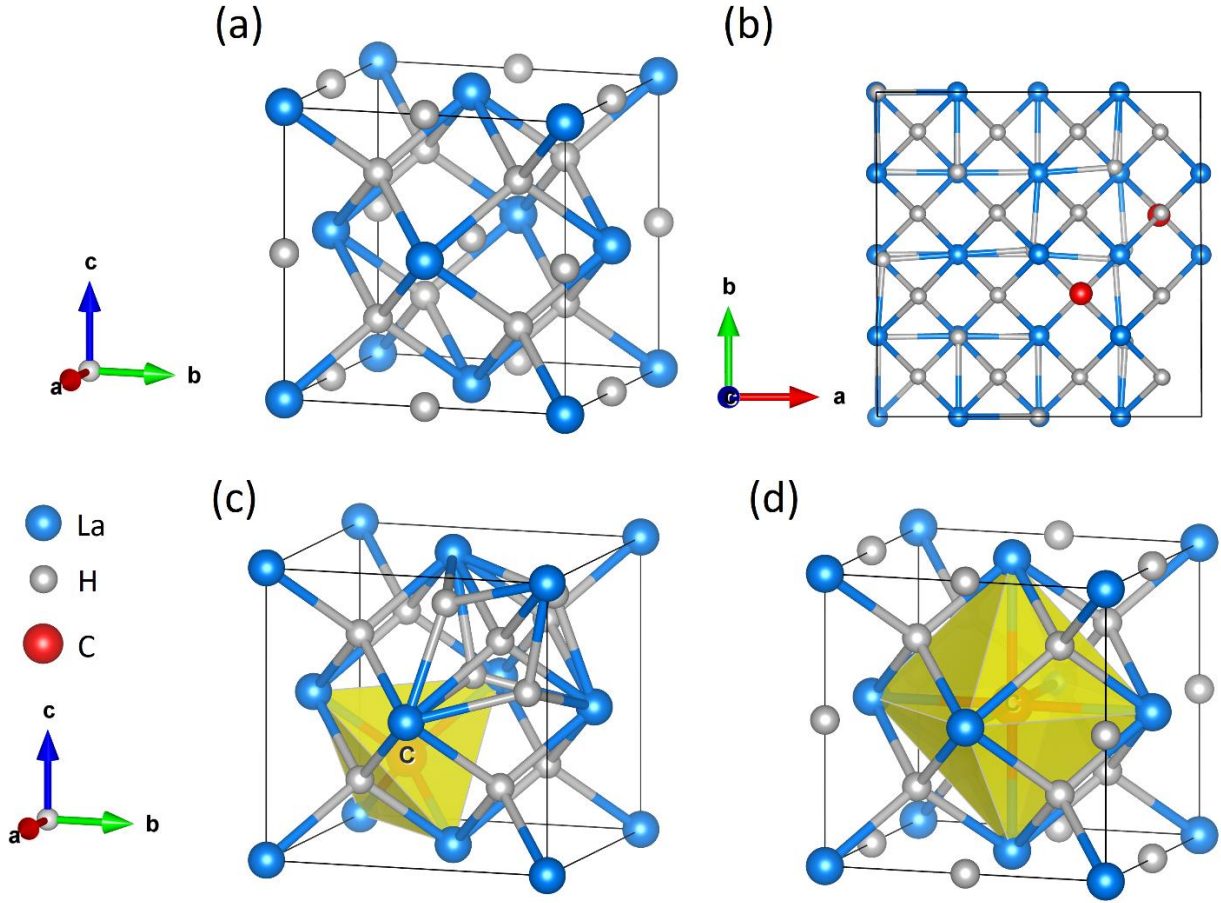

Figure 1. Relaxed LaH$_3$ crystal structure of (a) the pristine unit cell; (b) two C atoms doped in the 2 × 2 × 2 supercell; and (c,d) one C atom doped at the tetrahedral and octahedral sites of H in the unit cell, respectively. La, H, and C atoms are represented by the blue, grey, and red balls, respectively.

All calculations were performed using density functional theory (DFT) simulations, employing the Vienna Ab Initio Simulation Package (VASP) [28,29]. As it is bulk material, periodic boundary conditions have been applied in all 3 directions. The projector augmented wave method (PAW) with the Perdew-Burke-Ernzerhof (PBE) exchange-correlation functional within the generalized gradient approximation (GGA) [30] has been considered. Convergence tests for both k-point sampling of the Brillouin zone and cut-off energy are performed very carefully. For geometry optimization, the chosen threshold for Hellman-Feynman forces was $10^{-2}$ eV/A, whereas the energy convergence criterion was considered to be $10^{-5}$ eV/A. The valence electronic configurations of La, H, and C in the PAW pseudopotentials were considered as $5s^25p^65d^16s^2$, $1s^1$, and $2s^22p^2$, respectively [31]. The unit cell of LaH$_3$ contains 16 atoms (4 La and 12 H). For the k-point sampling of the Brillouin zone, a 13×13×13 (3×3×3) Monkhorst-Pack grid was employed in the geometry optimization, while a finer grid of 15 × 15 × 15 (5 × 5 × 5) k-points was used for the density of states (DOS) calculations in the unit cell (supercell) [32]. We used the state-of-the-art hybrid functional (HSE06) to confirm the induced magnetic moment results obtained with the PBE functional [33]. The cut-off energy of 500 eV was considered for the plane-wave basis set.

## 3. Results and discussion

### 3.1 Structure & Magnetic moment

LaH$_3$ is a nonmagnetic material, typically considered to have a cubic structure of D0$_3$ type (Strukturbericht designation) with the space group of $Fm\bar{3}m$ [34]. The optimized unit cell structure of pristine cubic LaH$_3$ is presented in Fig.1(a). The lattice constant obtained after relaxation is 5.56 Å, which matches well with the experimentally reported value of 5.6 Å [34,35]. The calculated minimum bond length between La-H is 2.40Å and 2.78 Å in the tetrahedral (HLa$_4$) and octahedral (HLa$_6$) coordination of the H atom, respectively. The calculated bond lengths are in good agreement with the previously reported values [36]. After substituting the C at the tetrahedral and octahedral sites of the H atom, a slight increase in bond length is found in both cases. Irrespective of the change in bond length, the structural symmetry remains preserved. Fig.1(b) shows the two C atoms substituted 2 ×2 ×2 supercell. Figs.1(c) and 1(d) represent the one C atom substituted optimized unit cell structures of LaH$_3$.

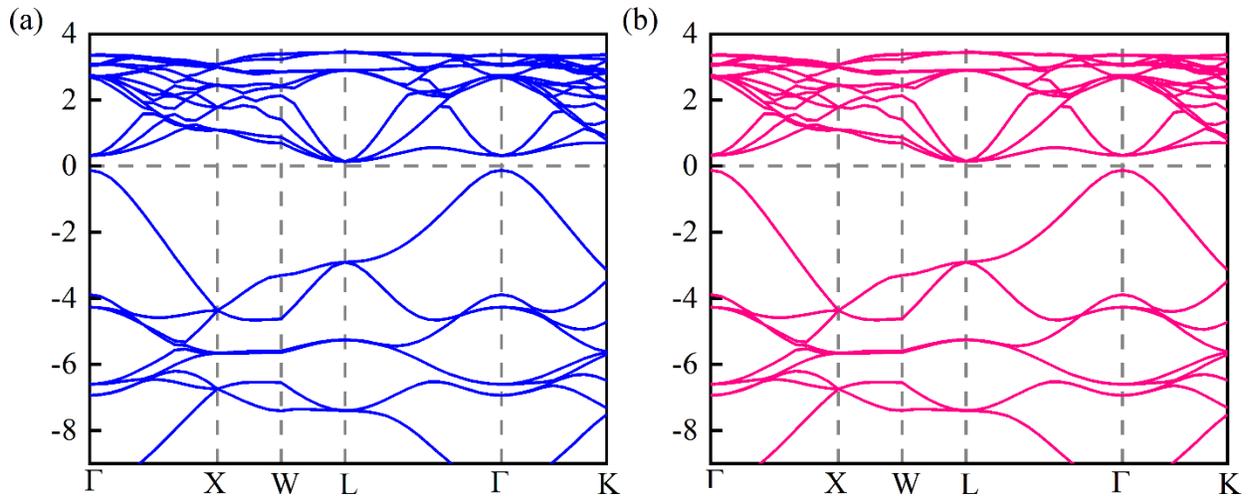

Figure 2. Electronic band structure of pristine LaH$_3$ for (a) up and (b) down spin states using HSE06 functional. The Fermi level is set at zero eV.

The NM LaH$_3$ has a wide range of applications in technology, hydrogen storage, rapidly diffusing materials in solids, etc. [37,38].Furthermore, hydrides are promising superconductor materials where T$_c$ (i.e., transition temperature) close to room temperature can be obtained at high pressure [39,40]. In the present work, NM cubic LaH$_3$ shows an indirect band gap of 0.26 eV using HSE06 functional. The corresponding band structure for spin-up and down states is shown in Fig.2. The substitution of the C atom at the octahedral and tetrahedral sites of H atom in NM LaH$_3$ introduce three additional electrons into the system. Normally, a magnetic moment of 3 µ$_B$ is expected from this type of substitution [41]. In order to acquire a thorough grasp of the induced magnetic moment and how it varies with impurity concentration, different configurations of the cubic unit cell of LaH$_3$ are considered. As shown in Table I, substitution of C at the tetrahedral (HLa4) and octahedral (HLa8) sites of H induce a magnetic moment of ~1.0 µ$_B$ corresponding to the impurity concentration of 1.04 at%. As shown in Table I, the substitution of C at the tetrahedral (HLa4) and octahedral (HLa8) sites of H induce a magnetic moment of ~1.0 µ$_B$ corresponding to the impurity concentration of 1.04 at %. With an increase in the impurity concentration, the magnetic moment is also increasing for both types of substitutions, reaching the value of 2.29 µ$_B$, except for the tetrahedral substitution case at 8.33 at%. This can be attributed to the appearance of relatively more spin-down states from the *p* orbital of C in the tetrahedral environment giving rise to a lesser net magnetic moment in this case compared to that of the octahedral environment. This can be further understood through the total density of states (TDOS) plots. Additionally, the octahedral site is found more favorable for C substitution due to it having a lower energy than the tetrahedral site. Further, corresponding to the octahedral substitution at the impurity concentration of 8.33 at%, a magnetic moment of 2.29 µ$_B$ is obtained using PBE functional (see Table I), whereas, with HSE06, a magnetic moment of 2.82 µ$_B$ is computed. The HSE06 functional is known to describe more accurately magnetic properties as compared to PBE due to the inclusion of Hartree-Fock part (25%) along with DFT. Therefore, a slightly different magnetic moment is obtained in the case of HSE06 functional as compared to PBE. The HSE06 confirms the induced magnetism in the system and supports our PBE functional-based results.

Table 1. Induced magnetic moment due to C substitution at tetrahedral and octahedral sites of H in LaH$_3$ using GGA.

| Impurity concentration (at%) | Magnetic Moments ($\mu_B$) | |
| --- | --- | --- |
| | Tetrahedral | octahedral |
| 1.04 | 0.99 | 0.97 |
| 2.08 | 1.65 | 1.57 |
| 4.16 | 2.27 | 2.11 |
| 8.33 | 1.00 | 2.29 |

### 3.2 DOS analysis

Next, in order to gain a better insight of the magnetic moment and the underlying mechanism for that, the TDOS and partial density of states (PDOS) of C-doped $LaH_3$ are examined. Figs.3(a) and 3(b) represent the TDOS corresponding to tetrahedral and octahedral C-doped $LaH_3$, respectively, at the doping concentration of 8.33 at%. The additional number of electrons that are introduced into the system due to C substitution causes spin splitting in the density of states, revealing the narrow, highly localized bands near the Fermi level. In the tetrahedral substitution case (see Fig.3(a)), due to spin splitting, overall, six spin channels appear in the TDOS (i.e., three up-spin channels and three down-spin channels) near the Fermi level. Out of six, two spin channels in the up-state and one spin channel in the down-state are completely occupied; the rest are unoccupied, giving a total magnetic moment of 1.0 $\mu_B$. On the other hand, in the octahedral substitution case (see Fig.3(b)), two spin channels are occupied completely, and one is partially occupied in the up-state; a magnetic moment of 2.29 $\mu_B$ is obtained in this substitution.

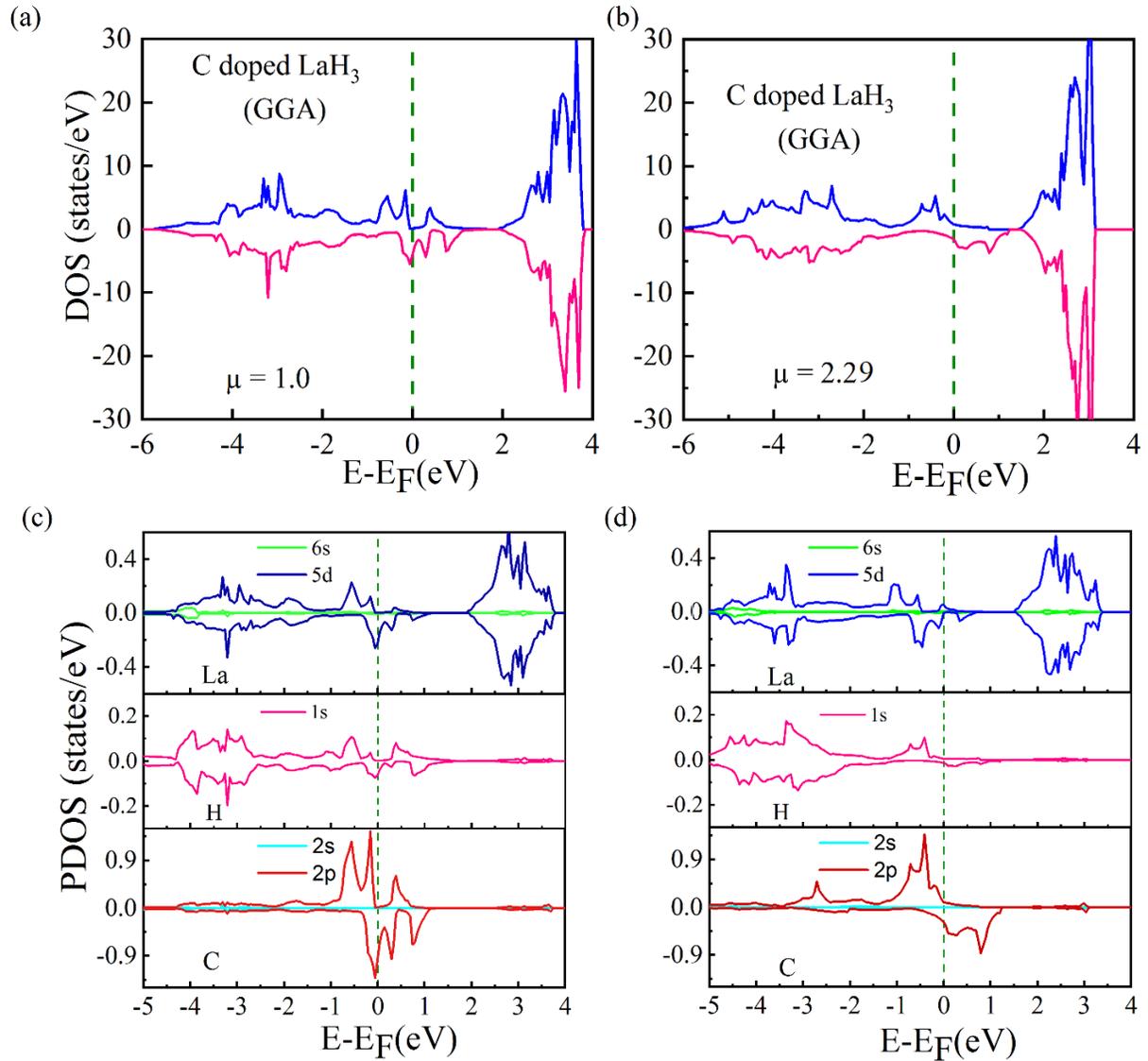

Figure 3. The total density of states (a,b) and partial density of states (c,d) corresponding to the C substitution at tetrahedral and octahedral positions of H, respectively, at the doping concentration of 8.33 at% using GGA. Blue color shows the spin-up and pink color shows the spin-down states in total density of states.

To understand more about spin splitting and individual orbital contributions to the magnetic moment, the partial density of states (PDOS) is also examined. Figs.3(c) and 3(d) display the PDOS of C-doped $LaH_3$ at tetrahedral and octahedral positions of H, respectively, corresponding to the doping concentration of 8.33 at%. We have calculated the PDOS for 2$s$ and 2$p$ orbitals of the C atom, 1$s$ orbital of the H atom, and 6$s$ and 5$d$ orbitals of the La atom. The 2$p$ orbital of the impurity C atom makes the maximum contribution to the induced magnetic moment, while nearby 5d and 1s orbitals of La and H atoms, respectively, also make small contributions. The *p-d* hybridization appears to be substantially stronger as compared to the *s-p* hybridization, based on the PDOS plots. Also, the impurity C, anion La, and cation H magnetically align in the same direction, supporting the FMs in the system. The impurity atom is found to contribute 59% out of 1.0 $\mu_B$ and 62% out of 2.29 $\mu_B$ in the tetrahedral and octahedral sites substitution, respectively. Compared to the tetrahedral substitution case (Fig.3(a)), the octahedral case (Fig.3(b)) has a higher number of up-spin channels relative to the down-channel. This can lead to a comparatively more net spin polarization in the octahedral case resulting in a higher magnetic moment.

### 3.3 Effect of Spin-Orbit Couplings

Further, we computed the magnetic moment by incorporating Spin-Orbit Coupling (SOC) in our calculation. We computed the magnetic moment (with SOC) corresponding to the minimum and maximum doping concentration (i.e., 1.04 at% and 8.33 at%). The addition of SOC does not change the computed magnetic moment considerably much; a small decrement in the magnetic moment of 0.01$\mu_B$ is found for both octahedral and tetrahedral substitution. Moreover, we found that with the inclusion of SOC, the electronic band structure changed a little bit in the tetrahedral substitution case, while in the octahedral substitution, the electronic band structure is not much affected with the inclusion of SOC (see Fig. 4).

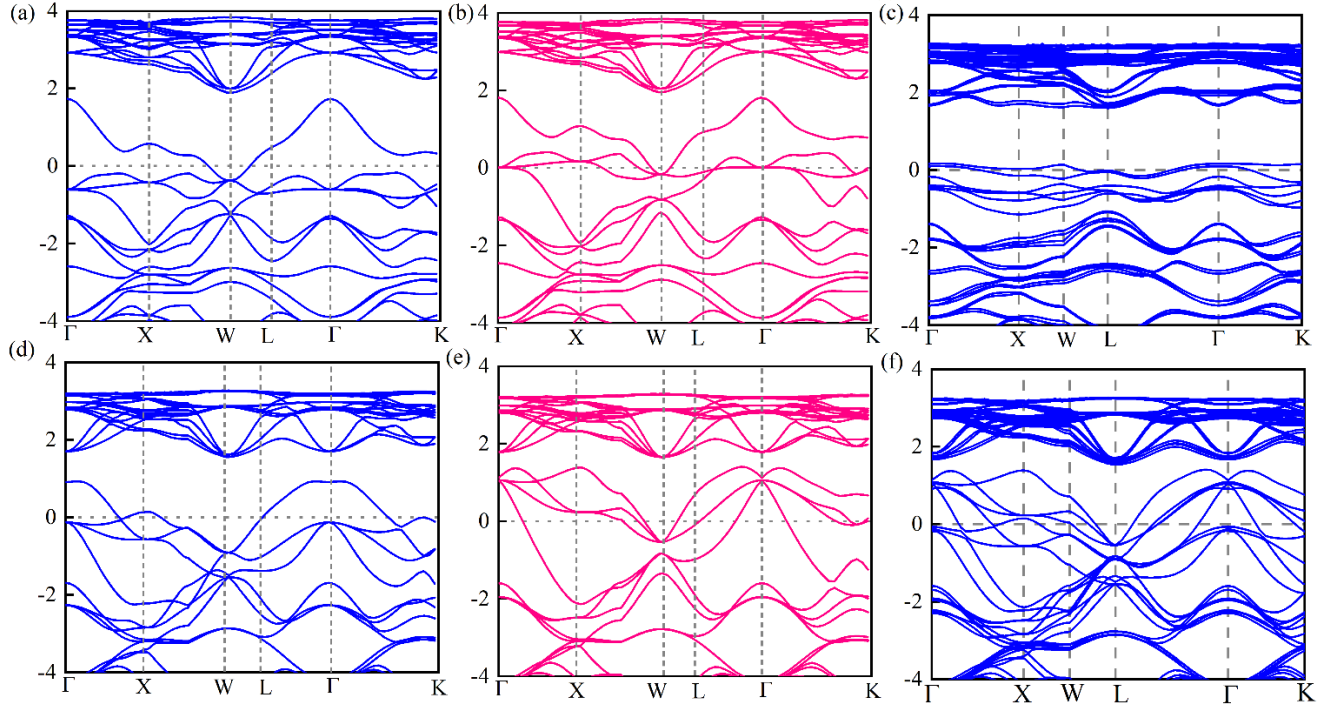

Figure 4. Electronic band structure of C-doped LaH3 at (a,b)tetrahedral and (d,e) octahedral sites for the spin-up and spin-down states, respectively, without SOC and (c,f) with SOC using GGA functional.

## 3.4 Spin Density Plot

In order to illustrate that the impurity atom contributes maximum in the magnetic moment, we also performed the spin density analysis. The total spatial extension of the real-space spin density isosurface gives a qualitative picture of the magnetic moment. Figs.5(a) and5(b) show the real-space effective spin density plots corresponding to C substitution at the tetrahedral and octahedral sites of H, obtained by subtracting the up ($\rho_\uparrow$) and down ($\rho_\downarrow$) spin-charge densities for the isovalue of 0.007e and for doping concentration of 8.33 at.%, respectively. The pink color represents the spin density isosurface, which is confined primarily around the impurity atom, verifying that the impurity atom contributes the most to the magnetic moment and a minimal contribution comes from nearby atoms. In the octahedral substitution case (Fig.5(b)), one can see the more spatial extension of spin density isosurface compared to the tetrahedral case (Fig.5(c))) around the impurity atom, supporting the appearance of a higher magnetic moment (2.29 μ$_B$) in the octahedral environment. This appearance of a higher magnetic moment is also supported by the PDOS analysis discussed earlier. We also analyze the spin density plot after substituting the single and double impurities in the 2×2×2 supercell with 128 atoms corresponding to the impurity concentration of 1.04 at% (Figs.5(c) and 5(d)), 2.04 at% for the isovalues of 0.001e and 0.005e, respectively. A similar pattern of spin density is obtained. This confirms that irrespective of dopant concentration, it is an impurity atom that induces magnetism in the system.

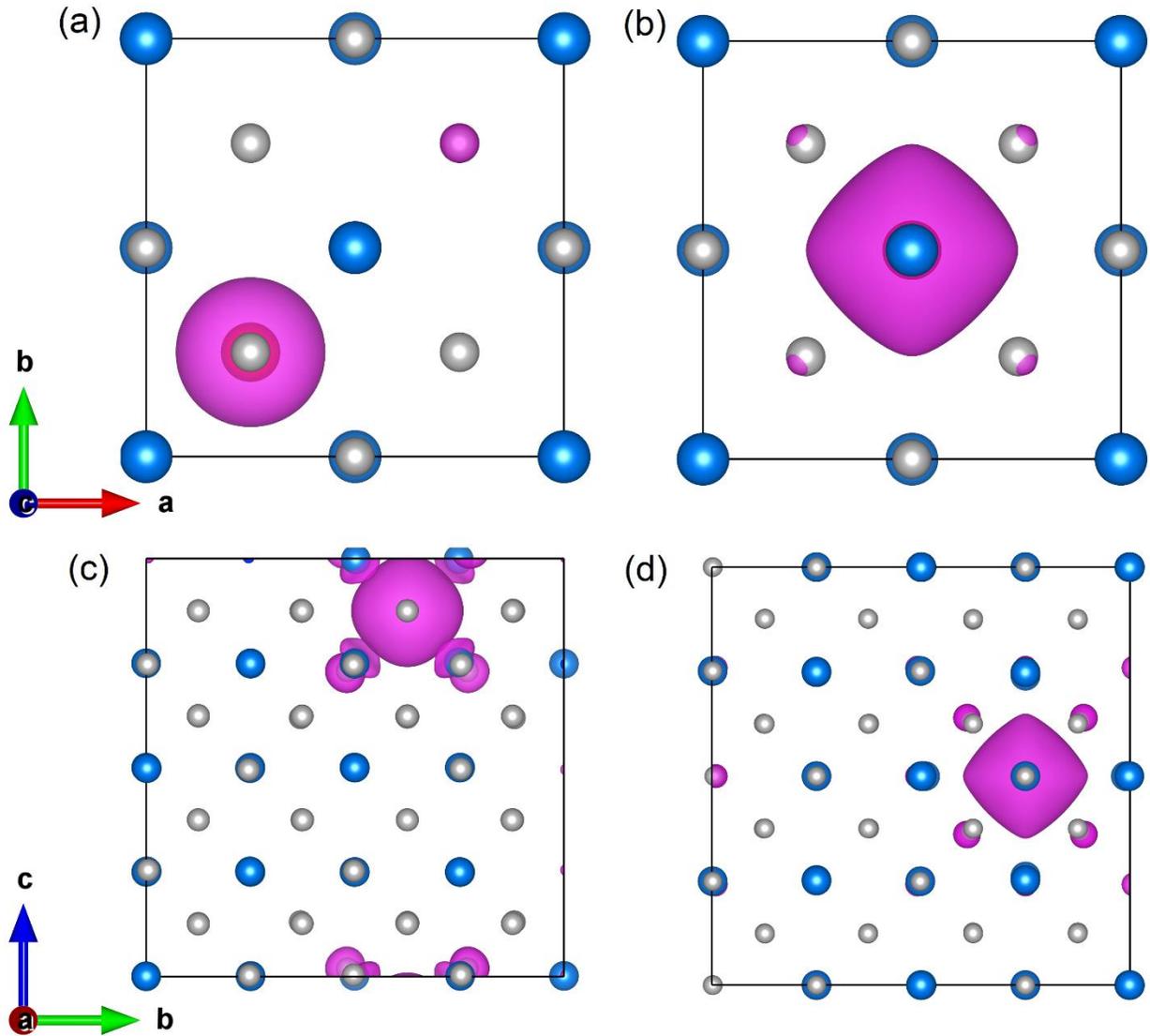

Figure 5. The real-space representation of the effective spin density plot of C-doped $LaH_3$ at tetrahedral and octahedral positions corresponding to the impurity concentration of (a,b) 8.33 at% and (c,d) 1.04 at% using GGA. The La and H atoms are shown in blue and grey colors, respectively. Purple color represents the spin density isosurface on the C atom.

**3.5 Computation of Curie temperature**

Using the DFT approach that has primarily been used for T = 0 K, we have discussed that the substitution of NM impurity in the NM LaH$_3$ induces magnetism. However, an essential difficulty for practical implementation is sustaining ferromagnetism at room temperature. The key feature of interest in the calculation of T$_c$ is to check whether the FMs survives at room temperature. Using mean-field approximation, T$_c$ has been computed employing the relation $T_c = \frac{2}{3} zR_{ij}$ where z denotes the number of the nearest neighbors [42]. The difference between the ferromagnetic (FM) and antiferromagnetic (AFM) energy (i.e., ΔE) is directly proportional to J, which further relies on the distance between the impurities atoms. Therefore, in order to evaluate T$_c$, it is important to study the magnetic interactions between the impurity atoms. This magnetic interaction is dependent on the distances between the two impurities. Therefore, we inserted two impurities in the system at different distances. Two impurity atoms are placed at four different distances: 2.40 Å, 3.93 Å, 4.6 Å, and 4.81 Å in the 2×2×2 supercell to evaluate T$_c$ as well as the actual magnetic ground state (see Fig.1(b)). We used the approach illustrated by Rahman et al. [17] and Zhao et al. [42] for the calculation of ΔE. While calculating the ΔE, all surroundings spins are taken into consideration in which spin from the periodic images of the supercell is also encountered, especially when a significant separation exists between the impurity atoms. Using the VASP code, the calculated values of exchange energy are reliable because the code itself considers the periodicity of the supercell. The variation of ΔE with impurity separations is presented in Fig.6(a). The FM coupling is found to be favorable at all considered separations in C-doped LaH$_3$. As the distance between the impurity atoms increases, ΔE begins to increase, attains a maximum value, and reaches zero at a distance of 4.81 Å. This indicates that the supercell under consideration should be large enough to account for all magnetic interactions. One advantage of the large range of magnetic interaction is that the system requires less impurity concentration to achieve room temperature FMs (RTF). With the information of ΔE, we computed exchange interaction and further T$_c$.

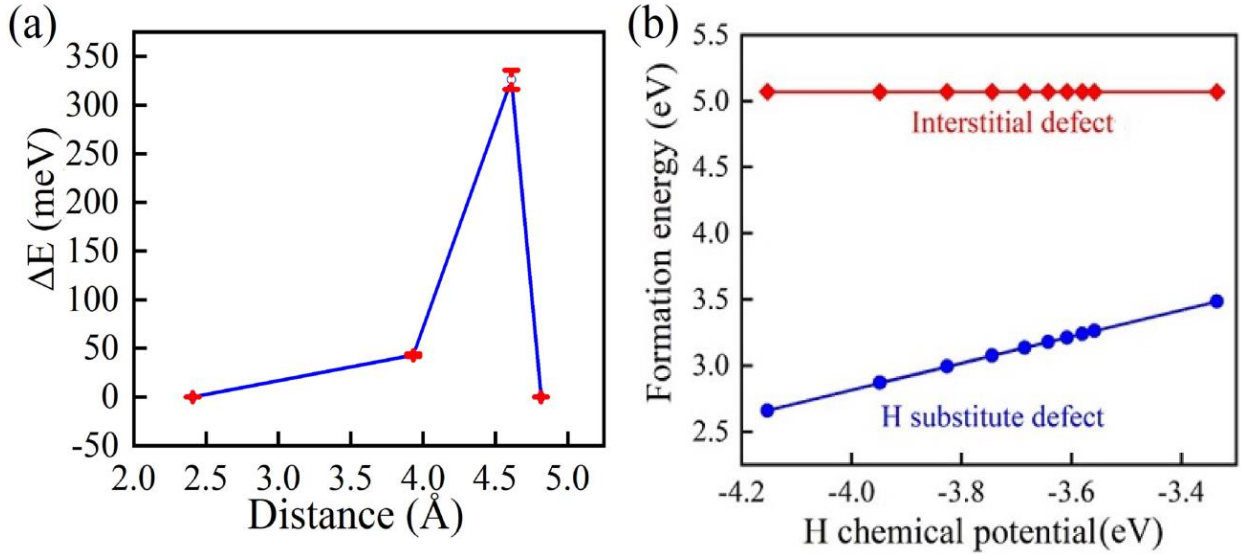

Figure 6. (a) Variation of ΔE with separation between impurities; red arrow indicates error bar; (b) Formation energy plot for H substitutional and interstitial defects with respect to H chemical potential.

The estimated Curie temperature using the mean-field approximation is *382 ± 12 K*. Although the mean-field approximation overestimates the Curie temperature, in higher doping (impurity) concentration limits, one can expect more accurate predictions from it. Kudrnovsky et al. [43] compared the experimentally computed results with the mean-filed approximation data in the case of Mn-doped GaAs and found a good agreement with the theoretical values in high impurity (Mn) concentration. In our study, the computed temperature is well above the room temperature, and we expect that the computed Curie temperature exists at least up to room temperature in C-doped LaH$_3$.

**3.6 Mechanism of Induced FM**

Further, we tried to understand the origin of induced magnetism. The existence of localized unpaired electrons in the 3*d* orbital of transition metals and 4*f* orbital of rare-earth metals are the main reasons behind magnetism in solids. Moreover, the properties of 2*p* electrons in C atoms are comparable to those of 3*d* orbitals of the Mn element of transition metal [44]. Also, as compared to 3*d* orbitals of the Mn atom, the 2*p* orbital causes larger spin-splitting energy. We are addressing here two possibilities that are responsible for spin polarization in the NM LaH$_3$ system: (i) localized nature of 2*p* orbital of the C atom, and (ii) it has large spin splitting energy. In addition, the relative bond strength of the cation-anion (or cation-impurity) is significant in introducing the local magnetic moment into the system. The 2*p* orbital of impurity atoms behaves like a localized orbital if the bond strength of the cation-impurity is weaker in comparison to the cation-anion bond strength that exists in the pristine system. On the other hand, if the bonding between the cation and impurity atoms is stronger, it leads to a strong hybridization; therefore, the system becomes non-spin polarized. When C is doped in nonmagnetic LaH$_3$, the La-C bond lengths are 2.47 Å and 2.79 Å, compared to La-H bond lengths of 2.40 Å and 2.78 Å in pure LaH$_3$. As a result of the weaker La-C bond, the impurity band becomes more localized, leading to the spin-polarized state. It is worth noting that the additional bands that appear near the Fermi level due to impurity substitution in the system do not confirm stable FMs. The defect band needs to be adjusted so that the energy gained from the exchange interaction should overcome the kinetic energy loss as per the band picture model in solid. Therefore, Stoner's condition, i.e., D(E$_F$)J>1, where D(E$_F$) denotes the TDOS near the Fermi level, and J is the Stoner parameter which represents the pairing energy (exchange energy) per pair, needs to be satisfied [27]. In our study, D(E$_F$) = 6.16 per eV when the separation between the two impurities is 4.61 Å, and the corresponding exchange energy interaction is 326 meV (Fig.6(a)). The product of D(E$_F$) and J came out to be greater than one in our study, satisfying the Stoner condition of FMs.

### 3.7 Practical Feasibility: Computation of Formation Energy

As discussed earlier, different kinds of defect configurations are possible corresponding to C substitution in the pristine cubic LaH$_3$ system such as hydrogen substitution defects, interstitial defects, and La substitution defects. For experimental feasibility of the system, formation energy of C-doped LaH$_3$ has been computed. The following formula has been used for the calculation of formation energy corresponding to different defects configurations:

$$E_{formation} = E_{defect} - E_{host} + \sum_k \Delta n_k \mu_k, \qquad (2)$$

where, E$_{defects}$ and E$_{host}$ represent the total energy per molecule obtained in the defective and pristine supercell, respectively. Moreover, Δn$_k$ denotes the number of atoms of k types that are removed (Δn$_k$< 0) or added (Δn$_k$> 0) in the supercell for creating defects in the system, and corresponding chemical potential is represented by μ$_k$. The chemical potentials give an estimation of the non-stoichiometry of the system, which is not achievable directly from the DFT calculations as a variety of factors such as pressures, growth circumstances, etc., affect it. Following the previous studies [45,46], we estimated the reasonable upper and lower bounds for the chemical potential and calculated the defects formation energy within these bounds. The chemical potential of La and H atoms is related to the energy per molecule of the pristine LaH$_3$ ($E_{LaH_3}$) system through the following relation:

$$\mu_{La}^{LaH_3} + 3\mu_H^{LaH_3} = E_{LaH_3}, \qquad (3)$$

where, $\mu_{La}^{LaH_3}$ and $\mu_H^{LaH_3}$ denote the chemical potentials of La and H atoms, respectively, in LaH$_3$. The stability bounds of LaH$_3$ with respect to the chemical potential of metallic lanthanum ($\mu_{La}^0$) and molecular hydrogen $\mu_H^0$ can be used to find out the maximum feasible range for $\mu_{La}^{LaH_3}$ and $\mu_H^{LaH_3}$, respectively. The below conditions (4) and (5) should be satisfied in order to avoid the formation of pure La and loss of hydrogen, respectively.

$$\mu_{La}^{LaH_3} < \mu_{La}^0, \qquad (4)$$

$$\mu_H^{LaH_3} < \mu_H^0. \qquad (5)$$

The formation energy of LaH$_3$ in terms of La and H chemical potentials is given by below relation:

$$E_{form}^{LaH_3} = \mu_{La}^{LaH_3} + 3\mu_H^{LaH_3} - (\mu_{La}^0 + 3\mu_H^0). \qquad (6)$$

By using the conditions (4), (5) and Eq. (6), the range for chemical potentials of La and H to evaluate the formation energies are obtained in LaH$_3$.

$$\mu_{La}^0 + E_{form}^{LaH_3} < \mu_{La}^{LaH_3} < \mu_{La}^0, \qquad (7)$$

$$\mu_H^0 + \frac{1}{3}E_{form}^{LaH_3} < \mu_H^{LaH_3} < \mu_H^0. \qquad (8)$$

Using the above formalism, formation energy as a function of H chemical potential has been calculated and presented in Fig.6(b). As in our study, the magnetic moment is obtained only in the hydrogen substitution defects; as a result, we mainly focus on studying the formation energy with respect to H chemical potential. The following relation can be used to vary the chemical potential of hydrogen by changing the experimental conditions such as temperature (T) and pressure (p):

$$\mu_H^0(T, P) = \mu_H^0(T, p^0) + \frac{1}{2} KT^* ln\left(\frac{p}{p^0}\right). \qquad (10)$$

Here, $p^0$ is equal to 1 atm and $k_B$ denotes the Boltzmann's constant. For favoring the hydrogen substitutional defects, $\mu_H^0$ should be tuned in such a way that formation energy in the case of hydrogen substitutional defects should be lower as compared to interstitial defects. In the formation energy plot (seeFig.6(b)), the blue and red colors represent the formation energy of H and interstitial defects with respect to H chemical potential, respectively. It is clear from the figure that formation energy for the H substitutional defects is lower than the interstitial defects for the complete range of H chemical potential. The obtained results suggest that the synthesis of C-doped LaH$_3$ is possible under experimental conditions because the formation energy of substitutional defects is lower in comparison with the interstitial defects.

## 4. Conclusion

In summary, we investigated that $d^0$ ferromagnetism can be instigated through anion substitution in nonmagnetic hydride. We showed that pure LaH$_3$ is nonmagnetic, but it exhibits ferromagnetism if we substitute the H atoms located on the octahedral and tetrahedral sites with C atoms. The localized behavior of the $2p$ states of C, along with significant exchange splitting energy, can be attributed to the origin of the induced spontaneous ferromagnetism. The induced ferromagnetism are found to increase with the increase in impurity concentration. For determining the true magnetic ground state of the system, two impurities were substituted in the 2×2×2 supercell, and by altering the distances between them, magnetic interactions were investigated. With a Curie temperature that is expected to be above the RT, ferromagnetic interactions were found to be strong enough to sustain the room-temperature ferromagnetism. The lower formation energy of hydrogen substitutional defects compared to interstitial defects validates the H substitutional doping in the system. We expect with the C substitution induces ferromagnetism in the system that can be used for the fabrication of spintronic devices in the near future.

**Data availability statement**

The data that support the findings of this study are available upon reasonable request from the authors.

**Acknowledgement**

P S acknowledges UGC, India, for the senior research fellowship [Grant No. 1330/(CSIR-UGC NET JUNE 2018)]. B C acknowledges Dr Nandini Garg, Dr T Sakuntala, Dr S M Yusuf, and Dr A K Mohanty for support and encouragement.